\documentclass[english, 11pt, a4paper]{article}

\usepackage{slashed}
\usepackage[utf8]{inputenc}
\usepackage{babel}
\usepackage{graphics}
\usepackage{graphicx}
\usepackage{texdraw}
\usepackage{epsfig}
\usepackage{float}
\usepackage{amsmath}
\usepackage{amssymb}
\usepackage{enumerate}
\usepackage[all]{xy}
\usepackage{fullpage}
\usepackage[small]{caption}
\usepackage{hyperref}

\newcommand{\be}{\begin{equation}}
\newcommand{\ee}{\end{equation}}

\begin{document}

\title{{\bf Interplay between $SU(N_f)$ chiral symmetry, $U(1)_A$ axial anomaly and massless
    bosons}}

\date{}
\author{Vicente ~Azcoiti \\
        Departamento de F{\'i}sica Te\'orica, Facultad de Ciencias, and \\
        Centro de Astropartículas y Física de Altas Energías (CAPA),\\ Universidad de Zaragoza,
        Pedro Cerbuna 9, 50009 Zaragoza, Spain\\}

\maketitle

\begin{abstract}

  The standard wisdom on the origin of massless bosons in the spectrum of a Quantum Field Theory
  $(QFT)$ describing the interaction of gauge fields coupled to matter fields is based on two
  well known features: gauge symmetry, and spontaneous symmetry breaking of continuous global
  symmetries.
  However we will show in this article how the topological properties, that originate the
  $U(1)_A$ axial anomaly in a $QFT$ which describes the interaction of fermion matter fields and
  gauge bosons, are the basis of an alternative mechanism to generate massless bosons in
  the chiral limit, if the non-abelian $SU(N_f)_A$ chiral symmetry is fulfilled in the vacuum. 
  We will also test our
  predictions with the results of a well known two-dimensional model, the two-flavour Schwinger
  model, which was analyzed by Coleman long ago, and will give a reliable answer to some of the 
  questions he asked himself on the spectrum of the model in the strong-coupling (chiral) limit.
  We will also analyze what are the expectations for the $U(N)$ gauge-fermion model in two
  dimensions, and will discuss on the impact of our results in the chirally symmetric 
  high temperature phase of $QCD$, which was present in the early universe, and is expected to be
  created in heavy-ion collision experiments. 
  To keep mathematical rigor we perform our calculations using a lattice regularization and
  Ginsparg-Wilson fermions.
\end{abstract}

\vfill\eject

\section{Introduction} \label{Introduction}

There are two well known mechanisms in Quantum Field Theory which allow us to understand the
existence of massless bosons in the spectrum of a given model of gauge fields coupled to matter
fields: gauge symmetry, and spontaneous symmetry
breaking of continuous global symmetries. The gauge symmetry is for instance responsible for the
photon not to have mass. On the other hand the spontaneous breaking of the $SU(2)_A$ chiral
symmetry in $QCD$ allows us to understand, via the Nambu-Goldstone theorem, why pions are so
light; indeed they would be massless if the up and down quark masses vanish.

However there are also some well known examples, as for instance two-flavour Quantum
Electrodynamics 
in ($1+1$) dimensions, in which chiral quasi-massless bosons appear in the spectrum of the model
near the chiral limit \cite{coleman}, and in which the explanation of this phenomenon escapes the
two aforementioned mechanisms to generate massless bosons. Hence it is worth wondering if this
happens because of some uninteresting peculiarities of two-dimensional models, or if there is a
deeper and general explanation for this phenomenon.

We want to show here how the topological properties of quantum field theories, which describe the
interaction of fermion matter fields and gauge bosons, and that exhibit $U(1)_A$ axial anomaly,
can be the basis of an alternative mechanism to generate massless bosons in the chiral limit.
More precisely we
will show, with the help of three distinct argumentation lines, that a gauge-fermion quantum
field theory, with $U(1)_A$ axial anomaly, and in which the chiral condensate vanishes in the
chiral limit, typically because of an exact non-abelian chiral symmetry, should exhibit a
divergent correlation length in the correlation function of the scalar condensate, in the
chiral limit. The non-anomalous Ward-Takahashi identities will tell us then that, in such a case,
also some pseudoscalar correlation functions should exhibit a divergent correlation length, 
associated to what would be the Nambu-Goldstone bosons if the non-abelian chiral symmetry were
spontaneously broken.

We will also test our predictions with the results of a well known two-dimensional model; 
the aforementioned two-flavour Schwinger model \cite{coleman}, and will discuss what are the
expectations for the $U(N)$ gauge-fermion model in two dimensions, the spectrum of which was
analyzed time ago in the large $N$ limit \cite{hooft}, and later on by means of bosonization
techniques \cite{affleck}, \cite{vento}. 

Some of the basic ideas here developed can be found in \cite{trece}, \cite{trecebis}. However, 
and in order to make this article self-contained, we will expose all of them in a detailed way.

Since the $Q=0$ topological sector will play a main role in our physical discussions, we will
devote section \ref{The relevance} 
to review some results concerning the relation between vacuum expectation
values of local and non-local operators computed in the $Q=0$ topological sector, with their
corresponding values in the full theory, taking into account the contribution of all topological
sectors, and will see how notwithstanding that the $Q=0$ sector breaks spontaneously the $U(1)_A$
axial symmetry, and shows a divergent pseudoscalar susceptibility in the chiral limit,
the associated pseudoscalar correlation length remains finite in this sector, and the
Nambu-Goldstone theorem is not fulfilled. To this end, we will analyze in this section the
one-flavour model, as well as the $N_f>1$ flavour theory, the latter in the case in which the
non-abelian axial symmetry is spontaneously broken, as it happens in the low temperature phase
of $QCD$. In section \ref{Two flavours} 
we will analyze what are
the physical expectations for the $N_f>1$ model when the non-abelian $SU(N_f)_A$ chiral symmetry
is fulfilled in the vacuum, and will show, with the help of three distinct argumentation
lines, how a theory which verifies the aforementioned properties should exhibit, in the chiral
limit, a divergent correlation length, and a rich spectrum of massless chiral bosons. 
Section \ref{Schwinger} is devoted to test the main prediction of this paper with well known
results of the two-flavour Schwinger model. In section \ref{un} we analyze our expectations for
the $U(N)$ gauge-fermion model in two-dimensions, and in the last section we report our
conclusions and discuss on the possible implications of our results in the high-temperature
chiral symmetry restored phase of $QCD$.

\section{Some relevant features of the $Q=0$ topological sector} \label{The relevance}

In this article we are interested in the analysis of some physical phenomena induced by the
topological properties of a fermion-gauge theory with $U(1)_A$ axial anomaly. In this analysis, 
the $Q=0$ topological sector will play an essential role, and this is the reason why we devote
this section to review some results concerning the relation between vacuum expectation values
of local and non-local operators computed in the $Q=0$ sector, with their corresponding
values in the full theory, where we take into account the contribution of all topological
sectors. In particular we will recall that the vacuum expectation value of local or intensive
operators computed in the $Q=0$ topological sector is equal, in the infinite volume limit, to their
corresponding value in the full theory. While this property is in general not true for non-local
operators, we will see later that there are exceptions. We will also show how, even if the
aforementioned property 
implies that the $U(1)_A$ symmetry is spontaneously broken in the $Q=0$ topological sector,
the Goldstone theorem is not realized because the divergence of the pseudoscalar susceptibility
does not come from a divergent correlation length \cite{trece}.

To begin let us write the continuum Euclidean Lagrangian for the most popular
gauge-fermion system with $U(1)_A$ axial anomaly, $QCD$ in four space-time dimensions. We want
to remark however that all the results reported in this paper apply to any gauge-fermion system
with $U(1)_A$ anomaly, and indeed in sections \ref{Schwinger} and \ref{un}  we will analyze
the Schwinger model
(two-dimensional $QED$) and the $U(N)$  model in two dimensions. The one-flavour $QCD$ Euclidean 
action in presence of a $\theta$-vacuum term reads as follows

\begin{equation}
  S = \int \mathrm{d}^4x \left\{\bar\psi\left(x\right)
  \left(\gamma_\mu D_\mu\left(x\right)+ m\right) \psi\left(x\right)
  + \frac{1}{4} F^a_{\mu\nu}\left(x\right)F^a_{\mu\nu}\left(x\right)
  + i\theta \frac{g^2}{64\pi^2} \epsilon_{\mu\nu\rho\sigma}
  F^a_{\mu\nu}\left(x\right)F^a_{\rho\sigma}\left(x\right)\right\}
  \label{eulagran}
\end{equation}
where $D_\mu(x)$ is the covariant derivative, and

\begin{equation}
  Q = \frac{g^2}{64\pi^2} \int d^4x\epsilon_{\mu\nu\rho\sigma}
  F^a_{\mu\nu}\left(x\right)F^a_{\rho\sigma}\left(x\right)
  \label{ftopcharg}
\end{equation}
is the topological charge of the gauge configuration, which is an integer number.

To give mathematical rigor to all developments along this paper we will avoid ultraviolet
divergences with the help of a lattice
regularization. We will also assume Ginsparg-Wilson (G-W) fermions \cite{Ginsparg},
the overlap fermions \cite{Neuberger1}, \cite{Neuberger2} being an explicit realization of them.
G-W fermions share 
with the continuum formulation all essential ingredients. Indeed G-W fermions
show an explicit $U(1)_A$ anomalous symmetry \cite{Luscher}, good chiral properties, a
quantized topological charge, and allow us to establish and exact index
theorem on the lattice \cite{Victor}. We recall here a few essential features of Ginsparg-Wilson
fermions which will be useful to understand the rest of the paper.

The lattice fermionic action for a massless G-W fermion can be written in a compact form as   

\begin{equation}
  S_F = \bar\psi D\psi
  \label{fa}
\end{equation}
where $D$, the Dirac-Ginsparg-Wilson operator, obeys the essential anticommutation equation

\begin{equation}
  D\gamma_5 + \gamma_5D = a D\gamma_5D
  \label{antic}
\end{equation}
where $a$ is the lattice spacing, and thus the right-hand side of (\ref{antic}) vanishes in the 
naive continuum limit, $a\rightarrow 0$.

It can be easily shown that action (\ref{fa}) is invariant under the following lattice $U(1)_A$
chiral rotation

\begin{equation}
\psi\rightarrow e^{i\alpha\gamma_5\left(I-aD\right)}\psi, \hskip 1cm \bar\psi\rightarrow
\bar\psi e^{i\alpha\gamma_5}
  \label{chirot}
\end{equation}
However the integration measure of Grassmann variables is not invariant, and the change of
variables (\ref{chirot}) induces a Jacobian

\begin{equation}
e^{-i 2\alpha \frac{a}{2} tr\left(\gamma_5 D\right)}
  \label{jacobian}
\end{equation}
where

\begin{equation}
\frac{a}{2} tr\left(\gamma_5 D\right) = n_- - n_+ = Q
  \label{topcharge}
\end{equation}
is an integer number, the difference between left-handed and right-handed zero modes, which can
be identified with the topological charge $Q$ of the gauge configuration. Thus equations
(\ref{jacobian}) and (\ref{topcharge}) show us how Ginsparg-Wilson fermions reproduce the
$U(1)_A$ axial anomaly. 

We can also add a symmetry breaking mass term, $m\bar\psi \left(1-\frac{a}{2}D\right)\psi$ to
action (\ref{fa}), so G–W fermions with mass are described by the fermion action 

\begin{equation}
  S_F = \bar\psi D\psi + m\bar\psi \left(1-\frac{a}{2}D\right)\psi
  \label{fam}
\end{equation}
and it can also be shown that the scalar and pseudoscalar condensates

\begin{equation}
S = \bar\psi \left(1-\frac{a}{2}D\right)\psi \hskip 1cm 
P = i\bar\psi \gamma_5\left(1-\frac{a}{2}D\right)\psi
\label{sapc}
\end{equation}
transform, under the chiral $U(1)_A$ rotations (\ref{chirot}), as a vector, just in the same way as
$\bar\psi\psi$ and $i\bar\psi\gamma_5\psi$ do in the continuum formulation.

The partition function of the $N_f$-flavour model in a finite lattice is the sum over all
topological sectors, $Q$, of the partition function in each topological sector times a 
$\theta$-phase factor,

\begin{equation}
Z = \sum_{Q} Z_Q e^{i\theta Q}
\label{zeta}
\end{equation}
where $Q$, which takes integer values, is bounded at finite volume by the number of
degrees of freedom. At large lattice volume the partition function should behave as

\begin{equation}
Z\left(\beta,m_f,\theta\right) = e^{-V E\left(\beta,m_f,\theta\right)}
\label{zetalarge}
\end{equation}
where $E\left(\beta,m_f,\theta\right)$ is the free energy density, $\beta$ the inverse 
gauge coupling, $m_f$ the $f$-flavour mass, 
and $V=V_s\times L_t$ the lattice volume in units of the lattice spacing. Moreover the 
partition function, and the mean value of any local or intensive operator $O$, as for instance
the scalar and pseudoscalar condensates, or any correlation function, in the $Q=0$ topological
sector, can be computed respectively as 

\begin{equation}
Z_{Q=0} = \frac{1}{2\pi}\int \mathrm{d}\theta Z(\beta,m_f,\theta)
\label{zq0}
\end{equation}

\begin{equation}
\left< O\right>^{Q=0} = \frac{\int \mathrm{d}\theta \left< O\right>_\theta Z(\beta,m_f,\theta)}
{\int \mathrm{d}\theta Z(\beta,m_f,\theta)}
\label{mascurioso}
\end{equation}
where $\left< O\right>_\theta$, which is the mean value of $O$ computed with the lattice
regularized integration 
measure (\ref{eulagran}), is a function of the inverse gauge coupling $\beta$, flavour masses 
$m_f$, and $\theta$, and it takes a finite value in the infinite lattice volume limit. 
Then, since the free energy density, as a 
function of $\theta$, has its absolute minimum at $\theta=0$ for non-vanishing fermion 
masses, the following relations hold in the infinite volume limit

\begin{equation}
E_{Q=0}\left(\beta,m_f\right) = E\left(\beta,m_f,\theta\right)_{\theta=0}
\label{eq0}
\end{equation}

\begin{equation}
\left< O\right>^{Q=0} = \left< O\right>_{\theta=0}.
\label{mascuriosob}
\end{equation}
where $E_{Q=0}\left(\beta,m_f\right)$ is the vacuum energy density of the $Q=0$ topological 
sector.\footnote{We want to notice that, as we will see later, equation (\ref{mascuriosob}) is in
  general wrong if some fermion mass vanishes.} As we will show below, equation
(\ref{mascuriosob}) is in general not true if $O$ is a non-local operator, while there are
exceptions to this rule.

We will devote the rest of this section to show that equation (\ref{mascuriosob}) is consistent
with the $U(1)_A$ axial anomaly. To this end, let us start with the analysis of the one-flavour
model at zero temperature.

In the one flavor model the only axial symmetry is an
anomalous $U(1)_A$ symmetry. The standard wisdom on the vacuum structure of
this model in the chiral limit is that it is unique at each given value
of $\theta$, the $\theta$-vacuum. Indeed, the only plausible reason to
have a degenerate vacuum in the chiral limit would be the spontaneous
breakdown of chiral symmetry, but since it is anomalous, actually there is
no symmetry. Furthermore, due to the chiral anomaly, the model shows a mass gap in the chiral
limit, and therefore all correlation lengths are finite in physical units. Since the model is
free from infrared divergences, 
the vacuum energy density can be expanded in powers of the
fermion mass $m_u$, treating the quark mass term as a perturbation \cite{Smilga}.
This expansion will be then an ordinary Taylor series

\begin{equation}
  E \left(\beta,m_u,\theta\right) = E_0\left(\beta\right)
  - \Sigma\left(\beta\right) m_u \cos\theta +O(m_u^2),
  \label{LS}
\end{equation}
giving rise to the following expansions for the scalar and pseudoscalar
condensates

\begin{equation}
  \left<S_u\right> =  -\Sigma\left(\beta\right) \cos\theta +O(m_u)
\end{equation}

\begin{equation}
  \left<P_u\right> =  -\Sigma\left(\beta\right) \sin\theta +O(m_u)
  \label{LSc}
\end{equation}
where $S_u$ and $P_u$ are the scalar and pseudoscalar condensates (\ref{sapc}) normalized by the
lattice volume. The topological
susceptibility $\chi_T$ is given, on the other hand, by the following expansion

\begin{equation}
  \chi_T = \Sigma\left(\beta\right) m_u \cos\theta +O(m_u^2)
\end{equation}

The resolution of the $U(1)_A$ problem is obvious if we set down the Ward-Takahashi identity 
which relates the pseudoscalar susceptibility 
$\chi_\eta = \sum_x \left<P_u\left(x\right)P_u\left(0\right)\right>$, the scalar condensate
$\left<S_u\right>$, and the topological susceptibility $\chi_T$

\begin{equation}
  \chi_\eta = -\frac{\left<S_u\right>}{m_u} -
  \frac{\chi_T}{m_u^2}.
  \label{Trans-II}
\end{equation}
Indeed the divergence in the chiral limit of the first term in the
right-hand side of (\ref{Trans-II}) is canceled by the divergence of the second term in
this equation, giving rise to a finite pseudoscalar susceptibility, and a finite non-vanishing
mass for the pseudoscalar $\eta$ boson.

Now we can apply equation (\ref{mascurioso}) to the computation of vacuum expectation values
of local operators, as the two-point pseudoscalar correlation function, but before that we want
to notice two relevant features of the $Q=0$ topological sector:

\begin{enumerate}
\item
  In the $Q=0$ sector the integration measure is invariant under global $U(1)_A$ chiral
  transformations
  because the full topological charge vanishes for any gauge configuration. This means that the
  global $U(1)_A$ axial symmetry is not anomalous in this sector.
  
\item
  If we apply equation (\ref{mascurioso}) to the computation of the vacuum expectation value
  of the scalar condensate, which is an intensive operator, we get that the $U(1)_A$ symmetry is
  spontaneously broken in the $Q=0$ sector because the chiral limit of the infinite volume
  limit of the scalar condensate, the limits taken in this order, does not vanish.
 
  \end{enumerate}

The two-point pseudoscalar correlation function
$\left<P_u\left(x\right)P_u\left(0\right)\right>$ is also an intensive operator, and equation
(\ref{mascurioso}) tell us that, in the infinite volume limit, and for $m_u\ne 0$, we can write

\begin{equation}
  \left<P_u\left(x\right)P_u\left(0\right)\right>^{Q=0} =
  \left<P_u\left(x\right)P_u\left(0\right)\right>_{\theta=0}.
  \label{mascurioso2}
\end{equation}
This equation implies that the mass of the pseudoscalar boson, $m_\eta$, which can be extracted
from the long distance behaviour of the two-point correlation function, computed in the $Q=0$
sector, is equal to the value we should get in the full theory, taking into account the
contribution of all topological sectors. On the other hand the topological susceptibility, 
$\chi_T$, vanishes in the $Q=0$ sector, and hence the Ward-Takahashi identity (\ref{Trans-II} )
in this sector reads as follows

\begin{equation}
  \chi_\eta^{Q=0} = -\frac{\left<S_u\right>^{Q=0}}{m_u}. 
  \label{wardq0}
\end{equation}
This identity gives us an expected result, the pseudoscalar susceptibility in the $Q=0$ sector
diverges in the chiral limit $m_u\rightarrow 0$ because the $U(1)_A$ symmetry is spontaneously
broken in this sector. Even if expected this is, however, a very surprising result because it
suggests that the pseudoscalar boson would be a Goldstone boson, and therefore its mass,
$m_\eta$, would vanish in the limit $m_u\rightarrow 0$.

The loophole to this paradoxical result is that in systems with a global
constraint, the divergence of the susceptibility does not necessarily implies
a divergent correlation length. The susceptibility is the infinite volume limit of the integral 
of the correlation function over all distances, in this order, and in systems with a global
constraint, the infinite volume limit 
and the space-integral of the correlation function do not necessarily commute. A very simple
and illustrative example is the Ising model at infinite temperature with an even number of spins, 
and vanishing full magnetization as global constraint \cite{trece}. In such a case
one has for the spin-spin correlation function

$$\hskip -3cm \left< s_i^2 \right>=1$$
$$\left< s_i s_j \right>= -\frac{1}{V-1}\hskip 0.5cm i\ne j$$
The integral of the infinite volume limit of the correlation function is equal to 1,
whereas the infinite volume limit of the integrated correlation function vanishes.
The correlation function has a contribution of order $1/V$, that violates cluster at
finite volume, and vanishes in
the infinite volume limit, but that gives a finite contribution to the
integrated correlation function. We will see in what follows how this is qualitatively what 
happens when computing the pseudoscalar correlation function in the $Q=0$ sector.

The $\left<P_u\left(x\right)P_u\left(0\right)\right>^{Q=0}$ correlation function at any finite
space-time volume $V$ verifies the following equation

\begin{equation}
  \left<P_u\left(x\right)P_u\left(0\right)\right>^{Q=0} = \frac{\int\mathrm{d}\theta 
    \left<P_u\left(x\right)P_u\left(0\right)\right>_{\theta} e^{-V E\left(\beta,m,\theta\right)}}
  {\int\mathrm{d}\theta e^{-V E\left(\beta,m,\theta\right)}}
  \label{menoscurioso}
\end{equation}
and we are interested not only in the infinite volume limit of this expression but also in the
$O\left(\frac{1}{V}\right)$ corrections.

On the other hand, it is standard wisdom that $QCD$ has no phase transition at $\theta=0$, and
hence we can expand the pseudoscalar correlation function in powers of the $\theta$ angle
as follows

\begin{equation}
  \left<P_u\left(x\right) P_u\left(0\right)\right>_{\theta} =
  \left<P_u\left(x\right) P_u\left(0\right)\right>_{\theta=0} + h(x, m_u) \theta^2
  + O(\theta^4)
  \label{expansion}
\end{equation}
where

\begin{equation}
  h(x, m_u) = \left\langle S_u\left(x\right) S_u\left(0\right)\right\rangle_{\theta=0} -
  \left\langle P_u\left(x\right) P_u\left(0\right)\right\rangle_{\theta=0} + O\left(m_u\right).
  \label{ache}
\end{equation}
$O\left(m_u\right)$ in (\ref{ache}) stays to indicate terms that vanish at least linearly with  
$m_u$ as $m_u\rightarrow 0$, in contrast with the first two terms in the right hand side
of (\ref{ache}) which take a non-vanishing value in the chiral limit.

The vacuum energy density can also be expanded in powers of $\theta$ as

\begin{equation}
  E \left(\beta,m_u,\theta\right) = E_{0}\left(\beta,m_u\right)
  - \frac{1} {2} \chi_T\left(\beta,m_u\right)\theta^2 +O(\theta^4)
  \label{expansion2}
\end{equation}

Taking into account equations (\ref{menoscurioso}), (\ref{expansion}) and(\ref{expansion2}), and
making an expansion around the saddle point solution we can write the following expansion in
powers of $\frac{1}{V}$  of the pseudoscalar correlation function in the zero-charge
topological sector

$$\left<P_u\left(x\right) P_u\left(0\right)\right>^{Q=0} =
\left<P_u\left(x\right) P_u\left(0\right)\right>_{\theta=0} +$$
\begin{equation}
  \frac{1}{V} \frac{\left\langle S_u\left(x\right) S_u\left(0\right)\right\rangle_{\theta=0} -
    \left\langle P_u\left(x\right) P_u\left(0\right)\right\rangle_{\theta=0}+ O(m_u)}{\chi_T} +
  + O\left(\frac{1}{V^2}\right)
  \label{saddlepoint}
\end{equation}
which shows, as in the simple Ising model case, a violation of
cluster at finite volume for the pseudoscalar correlation function in the zero-charge
topological sector, as follows from the fact that

\begin{equation}
  \lim_{|x|\rightarrow\infty} \left\langle S_u\left(x\right)
    S_u\left(0\right)\right\rangle_{\theta=0} = \Sigma^2.
\end{equation}
The cluster violating term is of the order of $\frac{1}{V}$, and because the topological
susceptibility $\chi_T = m_u\Sigma$ is linear in $m_u$, it is singular at $m_u=0$. It is just
this term who is responsible for the divergence of the pseudoscalar susceptibility in the
$Q=0$ sector in the chiral limit. However, 
in the infinite volume limit, the pseudoscalar correlation function
in the zero-charge topological sector and in the full theory at $\theta=0$ agree, as expected. 

In what concerns the pseudoscalar susceptibility, equations (\ref{menoscurioso}) and
(\ref{expansion}) allows us to relate this
quantity in the $Q=0$ sector and in the full theory as follows

\begin{equation}
  \chi_\eta^{Q=0} = \chi_{\eta_{\theta=0}} +
  \frac{\left(\left<S_u\right>_{\theta=0} - m_u \chi_{\eta_{\theta=0}}\right)^2}{\chi_T}
  \label{interesante}
\end{equation}
which shows explicitly how $\chi_\eta^{Q=0}$ diverges as $\frac{\Sigma}{m_u}$ when
$m_u\rightarrow 0$.

Summarizing we have shown that even if the $Q=0$ topological sector breaks spontaneously the
$U(1)_A$ axial symmetry to give account of the anomaly, the Goldstone theorem is not fulfilled
because the divergence of the pseudoscalar susceptibility does not come from a divergent
correlation length but from some peculiar features of the pseudoscalar correlation function
which can emerge in systems with global constraints.

The inclusion of more flavours does not change the qualitative results reported in this section 
when the $SU(N_f)$ chiral symmetry is spontaneously broken, as it happens in the
low temperature phase of $QCD$. The quantitative changes are essentially reduced to replace
the one-flavour scalar and pseudoscalar condensates by the flavour singlet scalar and
pseudoscalar condensates respectively, and the topological susceptibility $\chi_T$ by
$N_f^2\chi_T$ in equations (\ref{Trans-II}), (\ref{menoscurioso}-\ref{ache}) and
(\ref{saddlepoint}-\ref{interesante}). 

The case in which the $SU(N_f)$ chiral symmetry is fulfilled in
the vacuum will be discussed in detail in the next section.

\section{Two flavours and exact $SU(2)$ chiral symmetry} \label{Two flavours}

We will discuss in this section what are the physical expectations in a fermion-gauge theory
with two (or more flavours), exact $SU(2)$ chiral symmetry, and $U(1)_A$ axial anomaly.
In this discussion, the main ideas developed in the previous section will play an essential
role. We will see how a theory which
verifies the aforementioned properties should show, in the chiral limit, a divergent correlation
length, and a rich spectrum of massless chiral bosons. In section \ref{Vacuum energy density}
we will give,
under very general assumptions, a short demonstration of this result. Section \ref{The phase} 
contains a
qualitative but powerful argument supporting the results of section \ref{Vacuum energy density}, 
and in section \ref{Spectral density} we will
show how we can get the same qualitative result using general properties of the spectral density
of the Lee-Yang zeros of the partition function of the zero charge topological sector.

\subsection{Vacuum energy density of the $Q=0$ topological sector}
\label{Vacuum energy density}

As previously stated 
we consider a fermion-gauge model with two flavours, up and down, with masses $m_u$ and $m_d$, 
exact $SU(2)_A$ chiral symmetry, and $U(1)_A$ axial anomaly, as 
for instance the two-flavour Schwinger model or the high temperature phase of $QCD$. We will 
assume that the flavour-singlet scalar susceptibility
$\chi_\sigma\left(m_u, m_d\right)$, and hence also the flavour-singlet pseudoscalar 
susceptibility $\chi_\eta\left(m_u, m_d\right)$, take a finite value in the chiral limit, and 
will show that, in such a case, we get a quite surprising result: the scalar 
$\chi_\sigma\left(m_u, m_d\right)$ and
pseudoscalar $\chi_\eta\left(m_u, m_d\right)$ susceptibilities are equal in the chiral
limit, in contrast with what we would expect in a theory with two flavours and $U(1)_A$ anomaly.

To start the proof let us write the Euclidean fermion-gauge action (\ref{fam}) for the two-flavour
model, 

\begin{equation}
  S_F = m_u\bar\psi_u \left(1-\frac{a}{2}D\right)\psi_u +
        m_d\bar\psi_d \left(1-\frac{a}{2}D\right)\psi_d +
        \bar\psi_u D\psi_u + \bar\psi_d D\psi_d
  \label{conlag}
\end{equation}
where $D$ is the Dirac-Ginsparg-Wilson operator. 
This action can be written in a compact form as

\begin{equation}
  S_F = m_+\bar\psi \left(1-\frac{a}{2}D\right)\psi -
        m_-\bar\psi \left(1-\frac{a}{2}D\right)\tau_3\psi +
        \bar\psi D\psi
  \label{conlagcomp}
\end{equation}
where $m_+ = \frac{m_u+m_d}{2}$ and $m_- = \frac{m_d-m_u}{2}$. 
$\psi$ is a Grassmann field carrying site, 
Dirac, colour and flavour indices, and $\tau_3$ is the third Pauli matrix acting in flavour space. 

The vacuum energy density $E(m_+, m_-, \beta)$ of our model is a function of the quark masses, 
$m_+, m_-$, and the inverse gauge coupling $\beta$. Since we are assuming that the flavour-singlet
scalar susceptibility $\chi_\sigma$, and hence $\chi_\eta$, are finite in the chiral limit, and
because the pseudoscalar susceptibility $\chi_\eta$ is equal to the $\delta$-meson susceptibility
$\chi_\delta$ in this limit due to the exact $SU(2)_A$ axial symmetry, 
we can write a second order Taylor expansion for the free energy density as follows 

\begin{equation}
E\left(m_+, m_-, \beta\right) = \frac{1}{2} m_+^2 \chi_\sigma\left(\beta\right) + 
                                \frac{1}{2} m_-^2 \chi_\eta\left(\beta\right) + 
                                E_2\left(m_+, m_-, \beta\right) 
\label{taylor}
\end{equation}
where $E_2\left(m_+, m_-, \beta\right)$ verifies that 

$$
\lim_{m_+,m_-\rightarrow 0} \frac{E_2\left(m_+, m_-, \beta\right)}{m_+^2+m_-^2} = 0.
$$

We have shown in section \ref{The relevance}, equation (\ref{eq0}), that the vacuum energy 
density of the $Q=0$ topological sector is equal, in the thermodynamic limit, to the vacuum 
energy density in the full theory at $\theta=0$. Hence we can write

\begin{equation}
E_{Q=0}\left(m_+, m_-, \beta\right) = E\left(m_+, m_-, \beta\right)
\label{eq02}
\end{equation}

We can perform, in the $Q=0$ topological sector, an abelian axial rotation of the up-quark in the
path integral, with angle $\theta=\pi$, while leaving the down-quark unchanged. This variable 
change, the Jacobian of which is trivial in this sector, is equivalent to interchange $m_+$ and 
$m_-$, and so we get the following symmetry relation

\begin{equation}
E_{Q=0}\left(m_+, m_-, \beta\right) = E_{Q=0}\left(m_-, m_+, \beta\right)
\label{eq03}
\end{equation}

Equations (\ref{taylor}), (\ref{eq02}) and (\ref{eq03}) can only be verified if 
$\chi_\sigma\left(\beta\right) = \chi_\eta\left(\beta\right)$, 
and this concludes the proof.

This result tells us that a finite value of the flavour-singlet scalar susceptibility in the
chiral limit seems to be incompatible with the presence of the $U(1)_A$ axial anomaly in the
two-flavour model. In the next subsection we will give an argument pointing also to the divergence
of the flavour-singlet scalar susceptibility in the chiral limit for any $N_f\ge 2$.

\subsection{The phase diagram and the Landau approach} \label{The phase}

We have shown in section \ref{Vacuum energy density} that a fermion-gauge theory 
with $U(1)_A$ anomaly and exact $SU(2)$ chiral symmetry should exhibit a divergent correlation
length in the scalar sector in the chiral limit. 
In this section we want to give what is perhaps the strongest indication supporting this
result, which comes from a qualitative but powerful argument. To
this end we will explore the expected phase diagram of the model in the $Q=0$ topological sector
\cite{trecebis}, and will apply the Landau theory of phase transitions to it.

Since the $SU(2)$ chiral symmetry is assumed to be fulfilled in the vacuum, and the flavour
singlet scalar condensate is an order parameter for this symmetry, its vacuum expectation value 
$\left<S\right>=0$ vanishes in the limit in which the fermion mass $m\rightarrow 0$.
However, if we consider two non-degenerate fermion flavours, up and down, with masses $m_u$ and
$m_d$ respectively, and take the limit $m_u\rightarrow 0$ keeping $m_d\ne 0$ fixed, the up
condensate $S_u$ will reach a non-vanishing value

\begin{equation}
  \lim_{m_u\rightarrow 0} \left<S_u\right> = s_u\left(m_d\right) \ne 0
\label{ucond}
\end{equation}
because the $U(1)_u$ axial symmetry, which exhibits our model when $m_u=0$, is anomalous, and the
$SU(2)$ chiral symmetry, which would enforce the up condensate to be zero, is explicitly broken if
$m_d\ne 0$.

Obviously the same argument applies if we interchange $m_u$ and $m_d$, and we can therefore write
a equation symmetric to (\ref{ucond}) for the down condensate

\begin{equation}
  \lim_{m_d\rightarrow 0} \left<S_d\right> = s_d\left(m_u\right) \ne 0
\label{dcond}
\end{equation}
and since when $m_u, m_d\rightarrow 0$ the $SU(2)$ chiral symmetry is recovered and fulfilled in
the vacuum, we get

\begin{equation}
\lim_{m_d\rightarrow 0} s_u\left(m_d\right) = \lim_{m_u\rightarrow 0} s_d\left(m_u\right) = 0
\label{udcond}
\end{equation}

Let us consider now our model, with two non degenerate fermion flavours, restricted to the
$Q=0$ topological sector. As discussed in section \ref{The relevance}
the mean value of any local or intensive operator in 
the $Q=0$ topological sector will be equal, if we restrict ourselves to the region in which
both $m_u>0$, and $m_d>0$, to its mean value in the full theory in the infinite
lattice volume limit.\footnote{Since the two flavour model with $m_u<0$ and $m_d<0$ at
$\theta=0$ is equivalent to the same model with $m_u>0$ and $m_d>0$, this result is also true
if both $m_u<0$ and $m_d<0$.}We can hence apply this result to $\left<S_u\right>$ and
$\left<S_d\right>$ and write the following equations

$$\lim_{m_u\rightarrow 0} \left<S_u\right>^{Q=0} = s_u\left(m_d\right) \ne 0$$
\begin{equation}
  \lim_{m_d\rightarrow 0} \left<S_d\right>^{Q=0} = s_d\left(m_u\right) \ne 0
\label{dcondq0}
\end{equation}
In the $Q=0$ sector the $U(1)_u$ axial symmetry of our model at $m_u=0$, and the $U(1)_d$
symmetry at $m_d=0$ are good symmetries of the action because the Jacobian associated to a
chiral $U(1)_{u,d}$ transformation is the unit. Then equation (\ref{dcondq0}) tells us that
both, the
$U(1)_u$ symmetry at $m_u=0, m_d\ne 0$ and the $U(1)_d$ symmetry at $m_u\ne0, m_d=0$, are
spontaneously broken. This is not surprising at all since the present situation is similar 
to what happens in the one flavour model discussed in section \ref{The relevance},
and, as in that case, the
Goldstone theorem is not verified because the divergence of the pseudoscalar up or down
susceptibilities does not come from a divergent correlation length.

\begin{figure}[h!]
  \centerline{\includegraphics[scale=1.03]{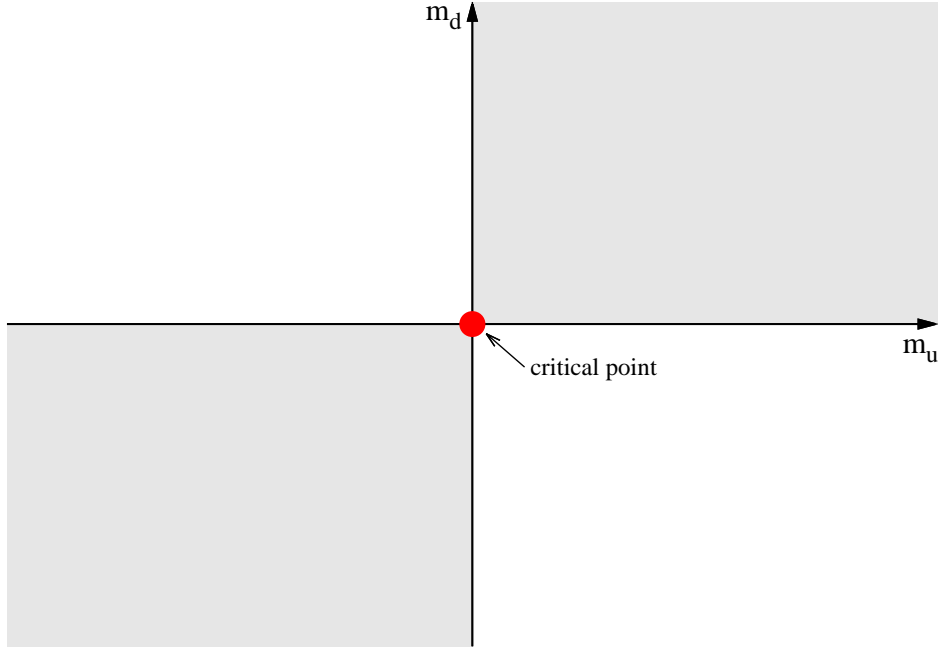}}
  \caption{Phase diagram of the two-flavour model in the $Q=0$ topological sector. The coordinate 
axis in the $(m_u,m_d)$ plane are first order phase transition lines. The origin of coordinates 
is the end point of all first order transition lines. The
vacuum energy density, its derivatives, and expectation values of local operators of
the two-flavour model at $\theta=0$ only agree with those of the $Q=0$ sector in the first
$(m_u>0,m_d>0)$ and third $(m_u<0,m_d<0)$ quadrants (the darkened areas).}
\end{figure}

Fig. 1 is a schematic representation of the phase diagram for the two-flavour model in the
$Q=0$ topological sector, and in the $(m_u, m_d)$ plane, which emerges from the previous
discussion. The two coordinate axis show first order phase transition lines. If we cross
perpendicularly the $m_d=0$ axis, the mean value of the down condensate jumps from $s_d(m_u)$
to $-s_d(m_u)$, and the same is true if we interchange up and down. All first order
transition lines end however at a common point, the origin of coordinates $m_u=m_d=0$,
where all condensates vanish because at this point we recover the $SU(2)$ chiral symmetry,
which is assumed to be also a symmetry of the vacuum. Notice that if the $SU(2)$ chiral
symmetry is spontaneously broken, as it happens for instance in the low temperature phase
of $QCD$, the phase diagram in the $(m_u, m_d)$ plane would be the same as that of Fig. 1
with the only exception that the origin of coordinates is not an end point.

Landau's theory of phase transitions predicts that the end point placed at the origin of
coordinates in the $(m_u, m_d)$ plane is a critical point, the scalar condensate should show
a non analytic dependence on the fermion masses $m_u$ and $m_d$ as we approach the critical
point, and hence the scalar susceptibility should diverge. But since the vacuum energy
density in the $Q=0$ topological sector matches the vacuum energy density in the full theory,
and therefore the same is true for the critical equation of state, Landau's theory
of phase transitions
predicts a non-analytic dependence of the flavour singlet scalar condensate on the
fermion mass, and a divergent correlation length in the chiral limit of our full theory, in which
we take into account the contribution of all topological sectors.

More precisely we can apply the Landau approach to analyze the critical behaviour around the two
first order transition lines in Fig.1 near the end point or critical point. In the analysis of the
$m_d=0$ transition line we consider $m_d$ as an external ''magnetic field''and $m_u$ as the
''temperature'', and vice versa for the analysis of the $m_u=0$ line. Then the standard Landau
approach tell us that the up and down condensates verify the two following equations of state

$$
m_u\left<S_u\right>^{-3} = -C_1 m_d \left<S_u\right>^{-2} + C_2
$$
\begin{equation}
m_d\left<S_d\right>^{-3} = -C_1 m_u \left<S_d\right>^{-2} + C_2
\label{eosud}
\end{equation}
where $C_1$ and $C_2$ are two positive constants. If we fix the ratio of the up and down masses
$\frac{m_u}{m_d}=\lambda$, the equations of state (\ref{eosud}) allow us to write the following
expansions for de up and down condensates

$$
  \left<S_u\right> = m_u^{\frac{1}{3}}\left(\left(\frac{1}{4C_2}\right)^\frac{1}{3} +
\frac{C_1}{3\left(2C_2^2\right)^\frac{1}{3}\lambda} m_u^\frac{1}{3} + \ldots\right)
$$
\begin{equation}
  \left<S_d\right> = m_d^{\frac{1}{3}}\left(\left(\frac{1}{4C_2}\right)^\frac{1}{3} +
\frac{C_1\lambda}{3\left(2C_2^2\right)^\frac{1}{3}} m_d^\frac{1}{3} + \ldots\right)
\label{expud}
\end{equation}
Equation
(\ref{expud}) shows explicitly the non analytical behaviour of the up and down condensates.
The flavour singlet scalar condensate scales as $m_u^\frac{1}{3} + m_d^\frac{1}{3}$ near the
critical point, and for the degenerate flavour case, $m_u=m_d=m$, we get

$$
\left<S\right> = \left<S_u\right> + \left<S_d\right> = 
\left(\frac{2}{C_2}\right)^\frac{1}{3} m^\frac{1}{3} + \ldots
$$
\begin{equation}
  \chi_\sigma\left(m\right) =  \frac{1}{3} \left(\frac{2}{C_2}\right)^\frac{1}{3} m^{-\frac{2}{3}}
  + \ldots
\label{landaucond}
\end{equation}
which shows explicitly the divergence of the flavour singlet scalar susceptibility in the
chiral limit.

The power low dependence of the scalar condensate (\ref{landaucond}) in the Landau approach
reproduces the mean field critical exponents. In general mean field exponents
are expected to be correct in four or higher dimensions. In lower dimensions the effect of
fluctuations can change the critical exponents, and this means that in these cases the Landau
approach give us a good qualitative description of the phase diagram but
fails in its quantitative predictions of critical exponents.

In general, and beyond the Landau approach, we can 
parameterize the critical behaviour of the flavour singlet scalar condensate for degenerate
flavours with a critical exponent $\delta>1$

\begin{equation}
  \left<S\right>_{m\rightarrow 0} \sim  m^{\frac{1}{\delta}}.
\label{criticaldelta}
\end{equation}
which gives us a divergent scalar susceptibility
$\chi_\sigma\left(m\right)\sim m^\frac{1-\delta}{\delta}$, and hence a massless scalar boson as
$m\rightarrow 0$. 

If on the other hand we write the Ward-Takahashi identity for the isotriplet of ''pions''
which follows from the $SU(2)_A$ non-anomalous chiral symmetry 

\begin{equation}
\chi_{\bar\pi}\left(m\right) = \frac{\left\langle S\right\rangle}{m},
\label{chipi2}
\end{equation}
we get that also $\chi_{\bar\pi}\left(m\right)$ diverges when $m\rightarrow 0$ as
$m^\frac{1-\delta}{\delta}$, and a rich
spectrum of massless bosons $(\sigma, \bar\pi)$ emerges in the chiral limit.

To conclude this section we would like to point out that the results reported here can be
generalized in a straightforward way to a number of flavours $N_f>2$.

\subsection{Spectral density of the Lee-Yang zeros of the partition function}
\label{Spectral density}

The results of sections \ref{Vacuum energy density} and \ref{The phase} 
have been obtained with the help of some general
properties of the vacuum energy density of the $Q=0$ topological sector. In view of the 
relevance of these results it is worth to explore some alternative way to corroborate it.  
In this section we will
show how we can get, using general properties of the spectral density of the Lee-Yang zeros of
the partition function of the zero charge topological sector, the same qualitative result in an
independent way.

We consider here a generic gauge-fermion model with $U(1)_A$ axial anomaly and two 
fermion flavours of equal mass, in which the $SU(2)$ chiral symmetry is fulfilled
in the ground state for massless fermions. 
Our starting assumption here, as in section \ref{Vacuum energy density},
is that the flavor singlet scalar
susceptibility, $\chi_\sigma\left(m\right)$, is a continuous function of the quark mass, $m$, at
$m=0$, in the full theory, taking into
account the contribution of all topological sectors. Under this assumption we will prove that
the flavor singlet scalar susceptibility, $\chi^{Q=0}_{\sigma}\left(m\right)$, in the $Q=0$ 
topological sector, is also a continuous function 
of the quark mass, $m$, at $m=0$; a result which together with the identities

$$\chi^{Q=0}_\sigma(m=0) =
\frac{1}{2}\chi_\sigma\left(m=0\right) + \frac{1}{2}\chi_\eta\left(m=0\right)$$

\begin{equation}
  \chi^{Q=0}_\sigma(m) = \chi_\sigma(m) \hskip1cm \forall m\ne 0,
\label{identi}
\end{equation}
will lead us to
the same paradoxical conclusion, $\chi_\sigma\left(m=0\right) = \chi_\eta\left(m=0\right)$,
obtained in section \ref{Vacuum energy density}.\footnote{The first
  of these identities can be easily derived from the $\theta$-dependence of the massless scalar
  susceptibility in the full two-flavour theory,
  $\chi_\sigma\left(\theta\right)=\cos^2\frac{\theta}
  {2}\chi_\sigma\left(m=0\right)+\sin^2\frac{\theta}{2}\chi_\eta\left(m=0\right)$.}

The zeros in the complex quark mass plane of the partition function of the zero charge
topological sector are distributed following several symmetry properties. If we denote with
$\mu$ de absolute value of a given zero, and with $\alpha$ its phase, the density of zeros
$\rho(\mu, \alpha)$ in the infinite lattice volume limit verifies the following symmetry relations

$$
\rho\left(\mu, \alpha\right) = \rho\left(\mu, -\alpha\right)
$$
\begin{equation}
\rho\left(\mu, \alpha\right) = \rho\left(\mu, \pi+\alpha\right)
\label{srelations}
\end{equation}
and then we can write the following expressions for the scalar condensate and the flavor singlet
scalar susceptibility, the last for massless fermions, 
in the $Q=0$ topological
sector

\begin{equation}
  \left< S\right>^{Q=0}\left(m\right) = -2 m\int  \mathrm{d}\alpha\int  \mathrm{d}\mu
  \frac{m^2 - \mu^2 \cos\left(2\alpha\right)}{m^4-2m^2\mu^2\cos\left(2\alpha\right) + \mu^4}
  \rho\left(\mu, \alpha\right)
\label{scondensate}
\end{equation}

\begin{equation}
  \chi^{Q=0}_{\sigma}\left(m=0\right) = 2 \int  \mathrm{d}\alpha\int  \mathrm{d}\mu
  \frac{\cos\left(2\alpha\right)}{\mu^2}
  \rho\left(\mu, \alpha\right)
\label{masslesssus}
\end{equation}
where $\alpha$ runs in the interval $(0,\pi)$, while is true that using the symmetry relations
(\ref{srelations}) the interval in $\alpha$ can be further reduced to $(0,\pi/2)$.

Since we are assuming that the flavor singlet scalar
susceptibility, $\chi_\sigma\left(m\right)$, takes a finite value when the quark mass goes to
zero, the scalar condensate at small fermion mass will be linear in the fermion mass, plus higher
order corrections. Furthermore, as discussed in previous sections, the scalar condensate and the
scalar susceptibility computed in the $Q=0$ topological sector agree, in the infinite lattice
volume limit, with the corresponding quantities computed in the full theory taking into account
the contribution of all topological sectors. Hence the chiral limit of
$\chi_{\sigma^{Q=0}}\left(m\right)$ can be computed as

\begin{equation}
  \lim_{m\rightarrow 0}\chi^{Q=0}_{\sigma}\left(m\right) =
  \lim_{m\rightarrow 0}\frac{\left< S\right>^{Q=0}\left(m\right)}{m} =
  -\lim_{m\rightarrow 0}2 \int  \mathrm{d}\alpha\int  \mathrm{d}\mu
  \frac{m^2 - \mu^2 \cos\left(2\alpha\right)}{m^4-2m^2\mu^2\cos\left(2\alpha\right) + \mu^4}
    \rho\left(\mu, \alpha\right)
\label{limssus}
\end{equation}
and the rest of this section will be devoted to show that the chiral limit (\ref{limssus}) is 
the massless flavor singlet scalar susceptibility
$\chi^{Q=0}_{\sigma}\left(m=0\right)$ (\ref{masslesssus}).

We should remark that the denominator in the right-hand side integral of (\ref{limssus}) vanishes
at $m^2 = \mu^2 e^{\pm i2\alpha}$, but since we assume that the model has no phase transitions
in the fermion mass $m$ near $m=0$, except at most at $m=0$, the zeros of the partition function
should stay at a finite distance of the real positive axis in the complex mass plane. Hence the
only candidate to be a singular point in the integrand of (\ref{limssus}) is $m=0, \mu=0$. This
means that if we split the $\mu$-integral into two regions, $\mu<\epsilon$, and $\mu>\epsilon$,
with $\epsilon\ll 1$, we can write

\begin{equation}
  \lim_{m\rightarrow 0}2 \int  \mathrm{d}\alpha\int_{\mu>\epsilon}  \mathrm{d}\mu
  \frac{m^2 - \mu^2 \cos\left(2\alpha\right)}{m^4-2m^2\mu^2\cos\left(2\alpha\right) + \mu^4}
  \rho\left(\mu, \alpha\right) = -2 \int  \mathrm{d}\alpha\int_{\mu>\epsilon}  \mathrm{d}\mu
  \frac{\cos\left(2\alpha\right)}{\mu^2}
  \rho\left(\mu, \alpha\right)
\label{multepsilon}
\end{equation}
and therefore we will concentrate on the chiral limit of the integral in the $\mu<\epsilon$ region.

Since we assume a finite massless scalar susceptibility $\chi_\sigma(m=0)$,
$\chi^{Q=0}_\sigma(m=0)=
\frac{1}{2}\chi_\sigma\left(m=0\right) + \frac{1}{2}\chi_\eta\left(m=0\right)$
will also be finite, and equation (\ref{masslesssus}) tells us that the
spectral density of zeros $\rho\left(\mu, \alpha\right)$ should vanish when $\mu\rightarrow 0$ 
fast enough in order to keep the $\mu=0$ singularity integrable. Hence we can parameterize the
behaviour of the spectral density of zeros near $\mu=0$ as

\begin{equation}
  \rho\left(\mu, \alpha\right)_{\mu\le\epsilon}\approx \mu^{p\left(\alpha\right)}
  f\left(\alpha\right)
\label{rholtepsilon}
\end{equation}
with $p\left(\alpha\right)>1$.\footnote{A value of $p\left(\alpha\right)\le 1$ could also give a
finite massless susceptibility (\ref{masslesssus}) if large 
cancellations when performing the $\alpha$-integral happen in a fine tuning way. For instance,
if we assume $p\left(\alpha\right)$ constant and less than 1, (\ref{masslesssus}) would be still
finite if $\int  \mathrm{d}\alpha \cos\left(2\alpha\right) f\left(\alpha\right) = 0$. However it
can be shown that, in such an unlikely case, the qualitative results obtained in this section do
not change.}

In order to compute the chiral limit of 

\begin{equation}
2 \int_0^\pi  \mathrm{d}\alpha \int_0^\epsilon  \mathrm{d}\mu
\frac{m^2 - \mu^2 \cos\left(2\alpha\right)}{m^4-2m^2\mu^2\cos\left(2\alpha\right) + \mu^4}
\rho\left(\mu, \alpha\right)
\label{vaya}
\end{equation}
we perform a change of variables and replace the spectral density of zeros in the previous
expression by its small $\mu$-value (\ref{rholtepsilon}), and so we get

$$2 \int_0^\pi  \mathrm{d}\alpha \int_0^\epsilon  \mathrm{d}\mu
\frac{m^2 - \mu^2 \cos\left(2\alpha\right)}{m^4-2m^2\mu^2\cos\left(2\alpha\right) + \mu^4}
\rho\left(\mu, \alpha\right)=$$

\begin{equation}
  \frac{2}{m} \int_0^\pi  \mathrm{d}\alpha \hskip 0.1cm m^{p\left(\alpha\right)}
  f\left(\alpha\right)
  \int^{\frac{\epsilon}{m}}_0  \mathrm{d}t \frac{1-t^2\cos\left(2\alpha\right)}
      {1-2t^2\cos\left(2\alpha\right)+t^4} t^{p\left(\alpha\right)}
\label{tintegral}
\end{equation}

It is easy to check that, for $p\left(\alpha\right)>1$,

\begin{equation}
  \lim_{m\rightarrow 0}\frac{2}{m} \int 
  \mathrm{d}\alpha\hskip 0.1cm m^{p\left(\alpha\right)}
  f\left(\alpha\right)
  \int^{\frac{\epsilon}{m}}_0  \mathrm{d}t \frac{1-t^2\cos\left(2\alpha\right)}
  {1-2t^2\cos\left(2\alpha\right)+t^4} t^{p\left(\alpha\right)} =
  -2 \int  \mathrm{d}\alpha \frac{f\left(\alpha\right)\cos\left(2\alpha\right)}
  {p\left(\alpha\right)-1}\epsilon^{p\left(\alpha\right)-1}
\label{alphagt1}
\end{equation}
and since the right-hand side of equation (\ref{alphagt1}) vanishes when
$\epsilon\rightarrow0$, we get that 

\begin{equation}
  \lim_{\epsilon\rightarrow0} \lim_{m\rightarrow 0} 2 \int 
  \mathrm{d}\alpha\int_0^\epsilon
  \mathrm{d}\mu
\frac{m^2 - \mu^2 \cos\left(2\alpha\right)}{m^4-2m^2\mu^2\cos\left(2\alpha\right) + \mu^4}
\rho\left(\mu, \alpha\right) = 0
\label{vaya2}
\end{equation}
a result which together equations (\ref{limssus}) and (\ref{multepsilon}) allow us to write

\begin{equation}
  \lim_{m\rightarrow 0}\chi^{Q=0}_{\sigma}\left(m\right) =
  \lim_{\epsilon\rightarrow0} 2 \int_0^\pi  \mathrm{d}\alpha\int_{\mu>\epsilon}  \mathrm{d}\mu
  \frac{\cos\left(2\alpha\right)}{\mu^2}
  \rho\left(\mu, \alpha\right)
\label{continuity}
\end{equation}
which tell us that the chiral limit of the flavour singlet scalar susceptibility in the $Q=0$
topological sector agrees with the massless scalar susceptibility in this sector, and therefore
the scalar susceptibility is a continuous function of the fermion mass, $m$, at $m=0$, in the
$Q=0$ sector. We should also notice that logarithmic violations 
to the power law behaviour of the spectral density $\rho\left(\mu, \alpha\right)$
(\ref{rholtepsilon}) do not change the previous qualitative result.

\section{The Schwinger model} \label{Schwinger}

The Schwinger model, or Quantum Electrodynamics in $(1+1)$-dimensions, is a good laboratory to
test the results reported in the previous sections. The model is confining \cite{kogut1},
exactly solvable at zero fermion mass, has non-trivial topology and shows explicitly the
$U(1)_A$ axial anomaly \cite{kogut2} through a non-vanishing value of the chiral condensate in
the chiral limit in the one-flavour case. Furthermore in the multi-flavour Schwinger model
the $SU(N_f)_A$ non-anomalous axial symmetry in the chiral limit is fulfilled in the vacuum,
and this property makes this model a perfect candidate to check the main conclusion of this
article, namely, the existence of light scalar and pseudoscalar bosons in the spectrum of the
model, the mass of which vanishes in the chiral limit.

The Euclidean continuum action is

\begin{equation}
  S = \int \mathrm{d}^2x \{\sum^{N_f}_{f=1}\bar\psi_f\left(x\right)
    \gamma_\mu\left(\partial_\mu +
  iA_\mu\left(x\right)\right)\psi_f\left(x\right) + 
  m \sum^{N_f}_{f=1}\bar\psi_f\left(x\right)\psi_f\left(x\right)
  + \frac{1}{4e^2} F^2_{\mu\nu}\left(x\right)\}
  \label{dos}
\end{equation}
where $m$ is the fermion mass and $e$ is the electric charge or gauge coupling, which has
the same dimension as $m$. $F_{\mu\nu}(x) = \partial_\mu A_\nu(x) - \partial_\nu A_\mu(x)$,
and $\gamma_\mu$ are $2\times 2$ matrices satisfying the algebra

\begin{equation}
  \{\gamma_\mu, \gamma_\nu\} = 2 g_{\mu\nu}
\end{equation}

This action is apparently invariant in the chiral limit under $SU(N_f)_A$ and $U(1)_A$ 
chiral transformations. However the $U(1)_A$-axial symmetry is broken at the quantum level
because of the axial anomaly. The divergence of the axial current is

\begin{equation}
  \partial_\mu J^A_\mu(x) = \frac{1}{2\pi} \epsilon_{\mu\nu} F_{\mu\nu}(x),
  \label{currentdivergence}
\end{equation}

\noindent
where $\epsilon_{\mu\nu}$ is an antisymmetric tensor, and hence does not vanish. 
The axial anomaly induces a topological $\theta$-term in the action of the form $i\theta Q$,
where 

\begin{equation}
  Q = \frac{1}{4\pi} \int \mathrm{d}^2x \epsilon_{\mu\nu} F_{\mu\nu}(x)
  \label{schtc}
\end{equation}
is the quantized topological charge.

The Schwinger model was analyzed years ago by Coleman \cite{coleman} computing some
quantitative properties of the theory in the continuum for both weak coupling,
$\frac{e}{m}\ll 1$, and strong coupling $\frac{e}{m}\gg 1$.

For the one-flavour model Coleman computed the particle spectrum of the model, which shows a mass
gap in the chiral limit, and conjectured 
the existence of a phase transition at $\theta=\pi$ and some intermediate fermion mass $m$
separating a weak coupling phase ($\frac{e}{m}\ll 1$) where the $Z_2$ symmetry of the
model at $\theta=\pi$ is spontaneously broken from a strong coupling phase
($\frac{e}{m}\gg 1$) where the $Z_2$ symmetry is realized in the vacuum.
A simple analysis of this model on the lattice also 
suggests that it should undergo a phase transition at some
intermediate fermion mass $m$ and $\theta=\pi$, even at finite lattice spacing.
Indeed, the lattice model is analytically solvable
in the infinite fermion mass limit (pure gauge two-dimensional electrodynamics with
topological term) \cite{puregauge}, and it is well known that the density
of topological charge approaches
a non-vanishing vacuum expectation value at $\theta=\pi$ for any value of the inverse
square gauge coupling $\beta$, showing spontaneous symmetry
breaking. On the other hand by expanding the vacuum energy density in powers of $m$, treating
the fermion mass as a perturbation, one gets for the vacuum expectation value
of the density of topological charge the following $\theta$-dependence

\begin{equation}
  \langle -i q\rangle = m\Sigma sin\theta + \frac{1}{2} m^2 \sin \left(2\theta\right)
  \left( \chi_\sigma - \chi_\eta\right) + \dots
  \label{qchiral}
\end{equation}

\noindent
where $\Sigma$ is the vacuum expectation value of the chiral condensate in the chiral limit and
at $\theta=0$ ($\Sigma = e^{\gamma_e}e/2\pi^{3/2}$ in the continuum limit), and $\chi_\eta$
and $\chi_\sigma$ are the pseudoscalar and scalar susceptibilities respectively. Equation
(\ref{qchiral}) shows how the $Z_2$ symmetry at $\theta=\pi$ is
realized order by order in the perturbative expansion of the topological charge in powers
of the fermion mass $m$. Therefore a critical point separating the large and small fermion mass
phases is expected, and this qualitative result has been recently confirmed by numerical
simulations of the Euclidean-lattice version of the model \cite{monos}. 

What is however more interesting for the content of this article is the Coleman analysis of the
two-flavour model. The theory has an internal $SU(2)_V\times SU(2)_A\times U(1)_V\times U(1)_A$
symmetry in the chiral limit, and the $ U(1)_A$ axial symmetry is anomalous. Since continuous
internal symmetries can not be spontaneously broken in a local field theory in two dimensions
\cite{coleman2}, the $SU(2)_A$ symmetry has to be fulfilled in the
vacuum, and the scalar condensate, which is an order parameter for this symmetry, will therefore 
vanish in the chiral limit, notwithstanding the chiral $U(1)_A$ anomaly. Hence the two-flavour
Schwinger model verifies all the conditions we assumed in section \ref{Two flavours}.

We summarize here the main Coleman's findings for the two-flavour model:

\begin{enumerate}
\item 
  For weak coupling, $\frac{e}{m}\ll 1$, the results on the particle spectrum are almost the
  same as for the massive Schwinger model.  
\item
  For strong coupling, $\frac{e}{m}\gg 1$, the low-energy effective theory depends only on one
  mass parameter, $m^\frac{2}{3} e^\frac{1}{3} \cos^\frac{2}{3}\frac{\theta}{2}$, the vacuum
  energy density will be then proportional to

  \begin{equation}
    E\left(m, e, \theta\right) \propto m^\frac{4}{3}
    e^\frac{2}{3}  \cos^\frac{4}{3} \frac{\theta}{2},
  \label{vacsch}
\end{equation}
and the chiral condensate at $\theta=0$ is therefore

  \begin{equation}
  \langle \bar\psi\psi\rangle \propto m^\frac{1}{3} e^\frac{2}{3}
  \label{chiralsch}
\end{equation}

\item
  The lightest particle in the theory is an
  isotriplet, and the next lightest is an isosinglet. The
  isosinglet/isotriplet mass ratio is $\sqrt 3$. If there are other
  stable particles in the model, they must be
  $O\left(\left[\frac{e}{m}\right]^\frac{2}{3}\right)$
  times heavier than these. The light boson mass, $M$, has a fractional power dependence on the
  fermion mass $m$:

  \begin{equation}
  M\propto m^\frac{2}{3} e^\frac{1}{3} \left(\cos\frac{\theta}{2}\right)^\frac{2}{3}
  \label{colmass}
  \end{equation}

\end{enumerate}

Many of these results have been corroborated by several authors both in the continuum
\cite{smilga2}, \cite{seiler}, \cite{james}, \cite{jac}, \cite{smilga3}, and using the lattice
approach \cite{gks}, \cite{gattr}. Coleman concluded his paper \cite{coleman} by asking some
questions concerning things he didn't understand, and we cite here two of them:

\begin{enumerate}
\item
  Why are the lightest particles in the theory a degenerate isotriplet?  
\item
  Why does the next-lightest particle has $I^{PG} = 0^{++}$, rather than $0^{--}$?

\end{enumerate}

We think that the results of section \ref{Two flavours} allow us to give a reliable answer
to these questions. The interplay between $U(1)_A$ anomaly and exact $SU(2)_A$ chiral symmetry
enforces the divergence of both, the flavour-singlet scalar susceptibility $\chi_\sigma$, and
the ''pion'' susceptibility $\chi_{\bar\pi}$ in the chiral limit. As discussed in section
\ref{The phase} both susceptibilities have the same fractional power dependence on the fermion
mass $m$,
$\chi_\sigma$,$\chi_{\bar\pi}\propto m^\frac{1-\delta}{\delta} e^\frac{\delta-1}{\delta}$, and
since

\begin{equation}
  {\chi_\sigma}_{m\rightarrow 0}\sim \frac{\mid\langle 0\mid \hat{O}_{\sigma}\mid
    \sigma\rangle\mid^2}{m_\sigma} \hskip 1cm
    {\chi_{\bar\pi}}_{m\rightarrow 0}\sim \frac{\mid\langle 0\mid \hat{O}_{\bar\pi}\mid \bar\pi
      \rangle\mid^2}
    {m_{\bar\pi}}
\label{lasdos}
\end{equation}
we expect the $\sigma$ and $\bar\pi$ masses have also the same dependence on the
fermion mass $m$,

\begin{equation}
    m_\sigma\propto \mid\langle 0\mid \hat{O}_{\sigma}\mid\sigma\rangle\mid^2
  m^\frac{\delta-1}{\delta} e^\frac{1-\delta}{\delta}, \hskip 1cm 
  m_{\bar\pi}\propto \mid\langle 0\mid \hat{O}_{\bar\pi}\mid
      \bar\pi\rangle\mid^2m^\frac{\delta-1}{\delta} e^\frac{1-\delta}{\delta}.
\label{msp}
\end{equation}
Coleman analysis predicts $\delta = 3$, which is the mean field critical exponent, and a finite
non-vanishing value for $\langle 0\mid \hat{O}_{\sigma}\mid\sigma\rangle$ and
$\langle 0\mid \hat{O}_{\bar\pi}\mid\bar\pi\rangle$ in the chiral limit.

Concerning the $\sigma$-meson pion mass ratio

$$\frac{m_\sigma}{m_{\bar\pi}}=\sqrt 3$$
reported by
Coleman in \cite{coleman} we find a discrepancy. The critical behaviour of the flavour-singlet
scalar condensate $\left<S\right>_{m\rightarrow 0} \sim  m^{\frac{1}{\delta}}$ 
beside the non-anomalous Ward-Takahashi identity (\ref{chipi2}) tells us that the ratio of the
pion and $\sigma$-meson susceptibilities will reach the value $\delta$ in the chiral limit

\begin{equation}
  \lim_{m\rightarrow 0} \frac{\chi_{\bar\pi} \left(m, e\right)}{\chi_\sigma \left(m, e\right)}
  = \delta
\label{nocol}
\end{equation}
and since the $SU(2)_A$ chiral symmetry is not spontaneously broken in the chiral limit, we
expect from (\ref{lasdos}) that  

\begin{equation}
  \lim_{m\rightarrow 0} \frac{\chi_{\bar\pi} \left(m, e\right)}{\chi_\sigma \left(m, e\right)}
  = \lim_{m\rightarrow 0} \frac{m_\sigma}{m_{\bar\pi}}
\label{nocol2}
\end{equation}
which, for $\delta=3$, give us the value 3 instead of $\sqrt3$ for the mass ratio. The origin
of this discrepancy may reside in the strong-coupling limit approximation made by Coleman in
\cite{coleman}. Indeed the bosonized two-flavour Schwinger model is a generalized Sine-Gordon
model which can not be solved in closed form, but in the strong coupling limit 
$\left(\frac{e}{m}\gg 1\right)$ approximation, the flavour-singlet pseudoscalar field 
is treated as a static field, and the model
is reduced to a special case of the standard Sine-Gordon model for the isotriplet pseudoscalar
field. Is inside the standard Sine-Gordon model where Coleman found that the $\sigma-\bar\pi$
mass ratio is $\sqrt 3$, but when going from the generalized Sine-Gordon model to the standard
Sine-Gordon model the structure of the mass term in the two-flavour Schwinger model is changed,
and hence the non-anomalous Ward-Takahashi identity (\ref{chipi2}), which depends on the
structure of the mass term, will also change. We want to notice, in this context, that the results
for the $\sigma-\bar\pi$ mass ratio of a numerical simulation of the two-flavour Schwinger
model with Kogut-Susskind fermions reported in \cite{gks} show a systematic deviation, at
large inverse gauge coupling $\beta=\frac{1}{e^2a^2}$ and small
values of the fermion mass, from the $\sqrt 3$ value, pointing to a larger value in the chiral
limit. This is however a rather old calculation, and an improvement of the results of \cite{gks}
could clarify this point.

We conclude this section by remarking that the results reported in section \ref{Two flavours} tell
us that the existence of quasi-massless chiral bosons in the spectrum of the two-flavour
Schwinger model near the chiral limit does not originates in some uninteresting peculiarities of
two-dimensional models but it should be a consequence of the interplay between exact non-abelian
chiral symmetry and $U(1)_A$ axial anomaly, and this is a picture that also holds for instance
in a much more
interesting case, the high temperature phase of four-dimensional $QCD$. What is a two-dimensional
peculiarity is the fact that in the chiral limit, when all fermion masses
vanish, these quasi-massless bosons become unstable and the low-energy spectrum of the model
reduces to a massless non-interacting boson, in accordance with Coleman's theorem \cite{coleman2}
which forbids the existence of massless interacting bosons in two dimensions.

\section{The $U(N)$  model in two dimensions} \label{un}

The analysis of the previous section on the multi-flavour Schwinger model applies also to the
$U(N)$ model in $(1+1)$ dimensions. The Euclidean continuum action is

\begin{equation}
  S = \int \mathrm{d}^2x \left\{\sum^{N_f}_{f=1}\bar\psi_f\left(x\right)
    \left(\gamma_\mu D_\mu\left(x\right)+ m_f\right) \psi_f\left(x\right)
  + \frac{1}{4e^2} F^a_{\mu\nu}\left(x\right) F^a_{\mu\nu}\left(x\right)\right\}
  \label{sun}
\end{equation}
where $D_\mu\left(x\right)$ is the covariant derivative, $\psi_f\left(x\right)$ a N-multiplet 
fermion field, $m_f$ the mass of flavour $f$, and the index $a$ runs from 1 to $N^2$. Since the
$U(1)$ electromagnetic field is also gauged in the $U(N)$ model, 
the $U(1)_A$ axial symmetry is, like in the Schwinger model, also anomalous in the $U(N)$ model in
$(1+1)$ dimensions. Furthermore the dimensionful coupling constant $e$ has mass dimensions, and the
model is also superrenormalizable.

In the one-flavour model we expect, as in the Schwinger model or in one-flavour four-dimensional
$QCD$, a mass gap in the spectrum in the chiral limit because of the
$U(1)_A$ axial anomaly. The spectrum of the $U(N)$ model
in $(1+1)$ dimensions was analyzed time ago in the large $N$ limit by 't Hooft \cite{hooft}, and
he found, in the one-flavour case, a spectrum of masses of the order of the gauge coupling, $e$,
plus a single mass which vanishes with the fermion mass. This massless boson appears in
the large $N$ limit because the effects of the $U(1)_A$ anomaly disappear at leading order in
this limit. Indeed at finite $N$ the one-flavour model shows a mass gap in the chiral
limit \cite{affleck}, as expected.

In what concerns the multi-flavour $U(N)$ model, 
we can apply the main conclusions of this paper. In the multi-flavour case the model has a
$SU(N_f)_A$ non-anomalous chiral symmetry and an anomalous $U(1)_A$  axial 
symmetry in the chiral limit. The $SU(N_f)_A$ chiral symmetry, as any continuous symmetry in
two dimensions, is not spontaneously broken \cite{coleman2}, and hence the scalar condensate 
$<S>=0$ vanishes in the chiral limit, notwithstanding the $U(1)_A$ anomaly. The results of section
\ref{Two flavours} lead us to conclude that the model should exhibit a divergent correlation
length in the chiral
limit, that together with the Ward-Takahashi identities analogous
to (\ref{Trans-II}) tells us that the
spectrum of the model should show $N_f^2$ quasi-massless chiral bosons near the chiral limit,
one of them scalar, and the other $N_f^2-1$ pseudoscalar.

\section{Conclusions and discussion}

The standard wisdom on the origin of massless bosons in the spectrum of a Quantum Field Theory 
describing the interaction of gauge fields coupled to matter fields is based on two well known
features: gauge symmetry, and spontaneous symmetry breaking of continuous symmetries. However, 
we have shown in this article that the topological properties, that originate the
$U(1)_A$ axial anomaly in a $QFT$ which describes the interaction of fermion matter fields and
gauge bosons, are the basis of an alternative mechanism to generate massless bosons in
the chiral limit, if the non-abelian $SU(N_f)_A$ chiral symmetry is fulfilled in the vacuum.
More precisely we have shown, with the help of three distinct argumentation
lines, that a gauge-fermion $QFT$, with $U(1)_A$ axial anomaly, and in which
the chiral condensate vanishes in the chiral limit, typically because of an exact non-abelian
chiral symmetry, should exhibit a divergent correlation length in the correlation function of
the scalar condensate, in the chiral limit. The non-anomalous Ward-Takahashi identities 
tell us then that, in such a case, also some pseudoscalar correlation functions should exhibit a
divergent correlation length, associated to what would be the Nambu-Goldstone bosons if the
non-abelian chiral symmetry were spontaneously broken.

The two-flavour Schwinger model, or Quantum Electrodynamics in two space-time dimensions,
is a good test-bed for our predictions. Indeed the Schwinger model shows a non-trivial topology, 
which induces the $U(1)_A$ axial anomaly. Moreover, in the two-flavour case, the non-abelian
$SU(2)_A$ chiral symmetry is fulfilled in the vacuum, as required by Coleman's theorem
\cite{coleman2} on the impossibility to break spontaneously continuous symmetries in two
dimensions.

The two-flavour Schwinger model was analyzed by Coleman long ago in \cite{coleman}, where 
he computed some quantitative properties of the theory in the continuum for both weak coupling,
$\frac{e}{m}\ll 1$, and strong coupling $\frac{e}{m}\gg 1$. In what concerns the strong-coupling
results, the main Coleman’s findings are qualitatively in agreement with our predictions.
The vacuum energy 
density (\ref{vacsch}), and the chiral condensate (\ref{chiralsch}) 
show a singular dependence on the fermion mass, $m$, in the chiral limit, and the flavour singlet
scalar susceptibility diverges when $m\rightarrow 0$. Moreover our results establish a 
reliable answer to some questions Coleman did himself \cite{coleman} concerning the following
two things he didn’t understand on the low-energy spectrum of the model.

\begin{enumerate}
\item
  Why are the lightest particles in the theory a degenerate isotriplet?  
\item
  Why does the next-lightest particle has $I^{PG} = 0^{++}$, rather than $0^{--}$?
\end{enumerate}
Indeed the interplay between the $U(1)_A$ anomaly and an exact $SU(2)_A$ chiral symmetry 
enforces the divergence of the flavour-singlet scalar susceptibility,
$\chi_\sigma\sim m^{\frac{1-\delta}{\delta}}$, $\delta>1$, in the
$m\rightarrow 0$ limit, and the non-anomalous Ward-Takahashi identity tell us that also the 
the ''pion'' susceptibility $\chi_{\bar\pi}\sim m^{\frac{1-\delta}{\delta}}$ diverges in the
chiral limit. The ratio value of these susceptibilities

$$\lim_{m\rightarrow 0}\frac{\chi_{\bar\pi}}{\chi_{\bar\sigma}}=\delta$$
implies on the other hand that the pion is lighter than the $\sigma$-meson.

The multi-flavour $U(N)$ model in $1+1$ dimensions is another test-bed for our predictions, and
we have analyzed this model in section \ref{un}. The results of this analysis are qualitatively
similar to those of the multi-flavour Schwinger model: the model spectrum should show
$N_f^2$ quasi-massless chiral bosons near the chiral limit, one of them scalar, and the other
$N_f^2-1$ pseudoscalar.

It is worth wondering if the reason for the rich spectrum of light chiral bosons near the chiral
limit found in the Schwinger and $U(N)$ models lies in some uninteresting peculiarities of
two-dimensional models, or if there is a deeper and general explanation for this phenomenon. 
We want to remark, concerning this, that our results reported in section \ref{Two flavours}
tell us that the existence of quasi-massless chiral bosons in the spectrum of these 
models near the chiral limit does not originates in some uninteresting peculiarities of
two-dimensional models but it should be a consequence of the interplay between an exact
non-abelian chiral symmetry and the $U(1)_A$ axial anomaly. What is a two-dimensional 
peculiarity is the fact that in the chiral limit, when all fermion masses
vanish, these quasi-massless bosons become unstable and the low-energy spectrum of the model
reduces to a massless non-interacting boson \cite{affleck}, \cite{vento}, in accordance with
Coleman's theorem \cite{coleman2} which forbids the existence of massless interacting bosons in
two dimensions.

In what concerns $QCD$, 
the analysis of the effects of the $U(1)_A$-axial anomaly in its high temperature phase,
in which the non-abelian chiral symmetry is restored in the ground state,
has aroused much interest in recent time, because of its relevance in axion phenomenology.
Moreover, the way in which the $U(1)_A$ anomaly manifests itself in the chiral
symmetry restored phase of $QCD$ at high temperature could be tested when probing the $QCD$ phase
transition in relativistic heavy ion collisions.

The first investigations on this subject started long time ago.
The idea that the chiral symmetry restored phase of two-flavor $QCD$ could be symmetric under
$U(2)\times U(2)$ rather than $SU(2)\times SU(2)$ was raised by Shuryak in 1994 \cite{shu}
based on an instanton liquid-model study. In 1996 Cohen \cite{cohen1} also got this result
formally from the QCD functional integral under some assumptions.
However immediately after several calculations questioning this result appeared
\cite{cuatro}-\cite{boy}. On the other hand a more recent analytic calculation of two-flavour
$QCD$ in the lattice, with overlap fermions, has shown \cite{diez} that the axial $U(1)_A$
anomaly becomes invisible in the scalar and pseudoscalar meson susceptibilities, suggesting again
that the effects of the anomaly disappear in the high temperature phase. However, as stated by
the authors of \cite{diez}, their result strongly relies on their assumption that the vacuum
expectation values of quark-mass independent observables, as the topological susceptibility, are
analytic functions of the square quark-mass, $m^2$, if the non-abelian chiral symmetry is
restored. Conversely, Coleman result for the topological susceptibility in the two-flavour Schwinger
model, which follows from equation (\ref{vacsch}), 

$$
\chi_T\propto m^\frac{4}{3}e^\frac{2}{3}
$$
shows explicitely a non-analytic quark-mass dependence, and casts serious doubts on the validity of 
this assumption.

The Dilute Instanton Gas Model \cite{diga1}-\cite{diga4} predicts on the other hand a topological
susceptibility for three light flavours, $\chi_T\sim \frac{1}{T^8}$, which decays with a power
law of the temperature at high $T$, and a recent lattice calculation \cite{guido} 
of the topological properties of full $QCD$ with physical quark masses, and temperatures around
$500 MeV$, gives as a result a small but non-vanishing topological susceptibility, although with
large error bars in the continuum limit extrapolations, suggesting that the effects of the
$U(1)_A$-axial anomaly still persist at these temperatures.

We can therefore do the reasonable hypothesis that the effects of the anomaly,
although diminished, still persist in 
the high temperature phase of $QCD$, and under such an assumption the main conclusions of this
paper should also apply to this phase. Taking into account recent lattice determination of the
light quark masses \cite{gilberto} ($m_u\simeq 2MeV$, $m_d\simeq 5MeV$, $m_s\simeq 94MeV$) we
can consider $QCD$ with two quasi-massless quarks as a good approach. Hence our results predict
a large value for the $\sigma$ and $\bar\pi$ meson susceptibilities, and a spectrum of
light $\sigma$ and $\bar\pi$ mesons at $T\gtrapprox T_c$, and the presence of these light scalar 
and pseudoscalar mesons in the chirally symmetric high temperature phase of $QCD$ could, on the 
other hand, significantly influence the dilepton and photon production observed in the particle 
spectrum \cite{dilepton} at heavy-ion collision experiments.

There are, on the other hand, two recent lattice calculations of mesonic screening masses in
two \cite{2f}, and three \cite{3f} flavour $QCD$ around, and above the critical temperature. The
reported results are not enough to allow a good check of our spectrum prediction. However, the
results of reference \cite{3f} show a small change of the pion screening-mass when crossing the
critical temperature, and a decreasing screening mass, at $T\gtrapprox T_c$, when going from the
$\bar us$ to the $\bar ud$ channel, compatible with a vanishing pion screening mass in the
chiral limit.

\section{Acknowledgments}

The author is grateful for many productive conversations with Giuseppe Di Carlo and Vicente Vento. 
This work was funded by Ministerio de
Economía y Competitividad under Grant No. FPA2015-65745-P (MINECO/FEDER).

\vfill
\eject

\vfill
\eject


\begin{thebibliography}{99}

\bibitem{coleman}
  S. Coleman, \textit{More about the Massive Schwinger Model},
  \textit{Ann. of Phys.} \textbf{101}, (1976) 239.

\bibitem{hooft}
  G. 't Hooft, \textit{A two-dimensional model for mesons},
  \textit{Nucl. Phys.} \textbf{B75}, (1974) 461.

\bibitem{affleck}
  I. Affleck, \textit{On the realization of chiral symmetry in (1+1) dimensions},
  \textit{Nucl. Phys.} \textbf{B265}[FS15], (1986) 448.

\bibitem{vento}
  A. Ferrando, V. Vento, \textit{The mesonic spectrum of bosonized $QCD_2$ in the chiral limit},
  \textit{Phys. Lett. B}\textbf{256} (1991) 503.

\bibitem{trece}
  V. Azcoiti, \textit{Topology in the $SU(N_f)$ chiral symmetry restored phase
    of unquenched $QCD$ and axion cosmology},
  \textit{Phys. Rev. D} \textbf{94}, (2016) 094505.

\bibitem{trecebis}
  V. Azcoiti, \textit{Topology in the $SU(N_f)$ chiral symmetry restored phase
    of unquenched $QCD$ and axion cosmology. II.}
  \textit{Phys. Rev. D} \textbf{96}, (2017) 014505.

\bibitem{Ginsparg}
  P.H. Ginsparg and K. G. Wilson, \textit{A remnant of chiral symmetry on the lattice},
  \textit{Phys. Rev. D} \textbf{25}, (1982) 2649.

\bibitem{Neuberger1}
  H. Neuberger, \textit{Exactly massless quarks on the lattice},
  \textit{Phys. Lett. B}\textbf{417} (1998) 141.

\bibitem{Neuberger2}
  H. Neuberger, \textit{More about exactly massless quarks on the lattice},
  \textit{Phys. Lett. B}\textbf{427} (1998) 353.
  
\bibitem{Luscher}
  M. Luscher, \textit{Exact chiral symmetry on the lattice and the Ginsparg-Wilson relation},
  \textit{Phys. Lett. B} \textbf{428} (1998) 342.

\bibitem{Victor}
  P. Hasenfratz, V. Laliena and F. Niedermayer, \textit{The index theorem in QCD with a finite
    cut-off},
  \textit{Phys. Lett. B} \textbf{427}, (1998) 125.

\bibitem{Smilga}
  H. Leutwyler and A. Smilga, \textit{Spectrum of Dirac operator and role of winding number in
    QCD},
  \textit{Phys. Rev. D} \textbf{46}, (1992) 5607.

\bibitem{kogut1}
  A. Casher, J. Kogut, L. Susskind, \textit{Vacuum polarization and the absence of free quarks},
  \textit{Phys. Rev.} \textbf{D10}, (1974) 732.

\bibitem{kogut2}
  J. Kogut, L. Susskind, \textit{How quark confinement solves the $\eta\rightarrow 3\pi$
    problem},
  \textit{Phys. Rev.} \textbf{D11}, (1975) 3594.

\bibitem{puregauge}
  U.J. Wiese, \textit{Numerical simulation of lattice $\theta$-vacua: The 2-d U(1) gauge theory
    as a test case},
  \textit{Nucl. Phys.} \textbf{B318}, (1989) 153.

\bibitem{monos}
  V. Azcoiti, E. Follana, E. Royo-Amondarain, G. Di Carlo, A. Vaquero Aviles-Casco,
  \textit{Massive Schwinger model at finite $\theta$},
  \textit{Phys. Rev. D} \textbf{97}, (2018) 014507.

\bibitem{coleman2}
  S. Coleman, \textit{There are no Goldstone Bosons in Two Dimensions},
  \textit{Comm. Math. Phys.} \textbf{31}, (1973) 259.

\bibitem{smilga2}
  A.V. Smilga, \textit{On the fermion condensate in the Schwinger model},
  \textit{Phys. Lett. B} \textbf{278}, (1992) 371.

\bibitem{seiler}
  C. Gattringer, E. Seiler, \textit{Functional integral approach to the N flavor Schwinger
    model},
  \textit{Ann. Phys.} \textbf{233}, (1994) 97.

\bibitem{james}
  J.E. Hetrick, Y. Hosotani, S. Iso, \textit{The massive multi-flavor Schwinger model},
  \textit{Phys. Lett. B} \textbf{350}, (1995) 92.

\bibitem{jac}
  A. Smilga, J.J.M. Verbaarschot, \textit{Scalar susceptibility in QCD and the multiflavor
    Schwinger model},
  \textit{Phys. Rev. D} \textbf{54}, (1996) 1087.

\bibitem{smilga3}
  A.V. Smilga, \textit{Critical amplitudes in two-dimensional theories},
  \textit{Phys. Rev. D} \textbf{55}, (1997) R443.

\bibitem{gks}
  C. Gutsfeld, H.A. Kastrup, K. Stergios, \textit{Mass spectrum and elastic scattering in the
    massive $SU(2)_f$ Schwinger model on the lattice},
  \textit{Nucl. Phys.} \textbf{B560}, (1999) 431.
  
\bibitem{gattr}
  C. Gattringer, I. Hip, C.B. Lang, \textit{The chiral limit of the two-flavor lattice
    Schwinger model with Wilson fermions},
  \textit{Phys. Lett. B} \textbf{466}, (1999) 287.

\bibitem{shu}
  E.V. Shuryak, \textit{Which chiral symmetry is restored in hot QCD?},
  \textit{Comments Nucl. Part. Phys.}\textbf{ 21}, (1994) 235.

\bibitem{cohen1}
  T. D. Cohen, \textit{QCD inequalities, the high temperature phase of QCD, and $U(1)_A$
    symmetry}
  \textit{Phys. Rev. D} \textbf{54}, (1996) R1867.

\bibitem{cuatro}
  C. Bernard, T. Blum, C. DeTar, S. Gottlieb, U. M. Heller,
  . E. Hetrick, K. Rummukainen, R. Sugar, D. Toussaint, and
  M. Wingate, \textit{Which Chiral Symmetry is Restored in High Temperature Quantum
    Chromodynamics?},
  \textit{Phys. Rev. Lett.} \textbf{78}, (1997) 598.

\bibitem{chandra}
  S. Chandrasekharan and N. H. Christ, \textit{Dirac spectrum, axial anomaly and the QCD chiral
    phase transition},
  \textit{ Nucl.Phys.Proc.Suppl.} \textbf{47}, (1996) 527.

\bibitem{kogut}
  J.B. Kogut, J.F. Lagae and D.K. Sinclair, \textit{Manifestations of the axial anomaly in
    finite temperature QCD},
  \textit{ Nucl.Phys.Proc.Suppl.} \textbf{53}, (1997) 269.

\bibitem{boy}
  G. Boyd, F. Karsch, E. Laermann and M. Oevers, \textit{Two flavor QCD phase transition},
  \textit{hep-lat/9607046} (1996).

\bibitem{diez}
  S. Aoki, H. Fukaya and Y. Taniguchi, \textit{Chiral symmetry restoration, the eigenvalue
    density of the Dirac operator, and the axial U(1) anomaly at finite temperature},
  \textit{Phys. Rev. D} \textbf{86}, (2012) 114512.
  
\bibitem{diga1}
  D.J. Gross, R.D. Pisarski and L.G. Yaffe, \textit{QCD and Instantons at Finite Temperature},
  \textit{Rev. Mod. Phys.} \textbf{53}, (1981) 43.

\bibitem{diga2}
  T.R. Morris, D.A. Ross and C.T. Sachrajda, \textit{Higher Order Quantum Corrections in the
    Presence of an Instanton Background Field},
  \textit{Nucl. Phys.} \textbf{B255}, (1985) 115.

\bibitem{diga3}
  T. Sch\"afer and E.V. Shuryak, \textit{Instantons in QCD},
  \textit{Rev. Mod. Phys.} \textbf{70}, (1998) 323.

\bibitem{diga4}
  A. Ringwald and F. Schrempp, \textit{Confronting instanton perturbation theory with QCD lattice
    results},
  \textit{Phys. Lett.} \textbf{B459}, (1999) 249.

\bibitem{guido}
  C. Bonati, M. D'Elia, G. Martinelli, F. Negro, F. Sanfilippo, A. Todaro, \textit{Topology in
    full QCD at high temperature: a multicanonical approach},
  \textit{JHEP} \textbf{11}, (2018) 170.

\bibitem{gilberto}
  FLAG working group of FLAVIANET, G. Colangelo et al., \textit{Review of lattice results
    concerning low energy particle-physics},
  \textit{Eur. Phys. J.} \textbf{C71}, (2011) 1695.

\bibitem{dilepton}
  R. Rapp, J. Wambach, \textit{Chiral Symmetry Restoration and Dileptons in Relativistic
    Heavy-Ion Collisions},
  \textit{Adv. Nucl. Phys.} \textbf{25}, (2000) 1.

\bibitem{2f}
  Bastian B. Brandt, Anthony Francis, Harvey B. Meyer, Owe Philipsen, Daniel Robainad and
  Hartmut Wittig, \textit{On the strength of the $U_A(1)$ anomaly at the chiral phase transition
  in $N_f = 2$ $QCD$},
  \textit{JHEP} \textbf{12}, (2016) 158.

\bibitem{3f}
  A. Bazavov et al., (HotQCD collaboration), \textit{Meson Screening Masses in (2+1)-Flavor
  $QCD$},
  \textit{arXiv:1908.09552}.

  
\end{thebibliography}
\end{document}